\documentclass[aps,prl,twocolumn,a4paper,english]{revtex4-2}

\usepackage{newtxtext,newtxmath}
\usepackage{color}
\usepackage{amsmath}
\usepackage{graphicx,float}  
\usepackage{epsfig, subfigure}
\usepackage{mathrsfs}
\usepackage{bm}        
\usepackage{amsfonts}
\usepackage{amssymb}
\usepackage{mathrsfs}
\graphicspath{{figs/}}
\usepackage{natbib}

\usepackage{epsfig,subfigure}
\usepackage{multirow}
\usepackage{caption}
\usepackage{verbatim}
\usepackage{gensymb}

\begin{document}


\title{Origin of hysteresis in shock wave reflection}


\author{Yan-Chao Hu, Zhi-Gong Tang, Yan-Guang Yang, Wen-Feng Zhou, Zhao-Hu Qin}
\affiliation{China Aerodynamics Research and Development Center;\\Peking University}


\date{\today}

\begin{abstract}
We report the mechanism of the hysteresis in the transition between Regular and Mach reflections. A new discovery is that, the hysteresis loop is in fact the projection of a higher dimensional path, i.e. the valley lines in the surface of dissipation, of which minimal values correspond to stable reflection configurations. Since the saddle-nodes bifurcate the valleies of the surface, they are actually the transition points of the two reflections. Furthermore, the predicted reflection configurations agree well with the experimental and numerical results, which is a validation of this theory.
\end{abstract}


\maketitle


\par Hysteresis is a general property of systems with two or more possible steady states, where hysteresis loops always emerge as external parameters vary continuously. A canonical example is the system described by the ferromagnetism theory \cite{MichaelPlischke2006Equilibrium}, where cyclical variation of the magnetic field intensity $H$ induces a hysteresis loop of the magnetization $\kappa$. $H$ is the external parameter of the system, and $\kappa$ is the order parameter, proposed by Landau \cite{goldenfeld2018lectures},  manifesting the ferromagnet state. Other systems such as liquid–solid phase transitions \cite{WilkinsonPositron,xu2006large}, laminar-turbulent transitions \cite{HofFinite,BenCritical,AvilaThe,LemoultDirected} and Bose–Einstein condensation \cite{MuellerSuperfluidity,diakonov2002loop,MorschDynamics,EckelHysteresis} all possess this property. 
\par
\begin{figure}
	\subfigure[\label{subfig:schlieren}]{\includegraphics[width = 1.0\columnwidth]{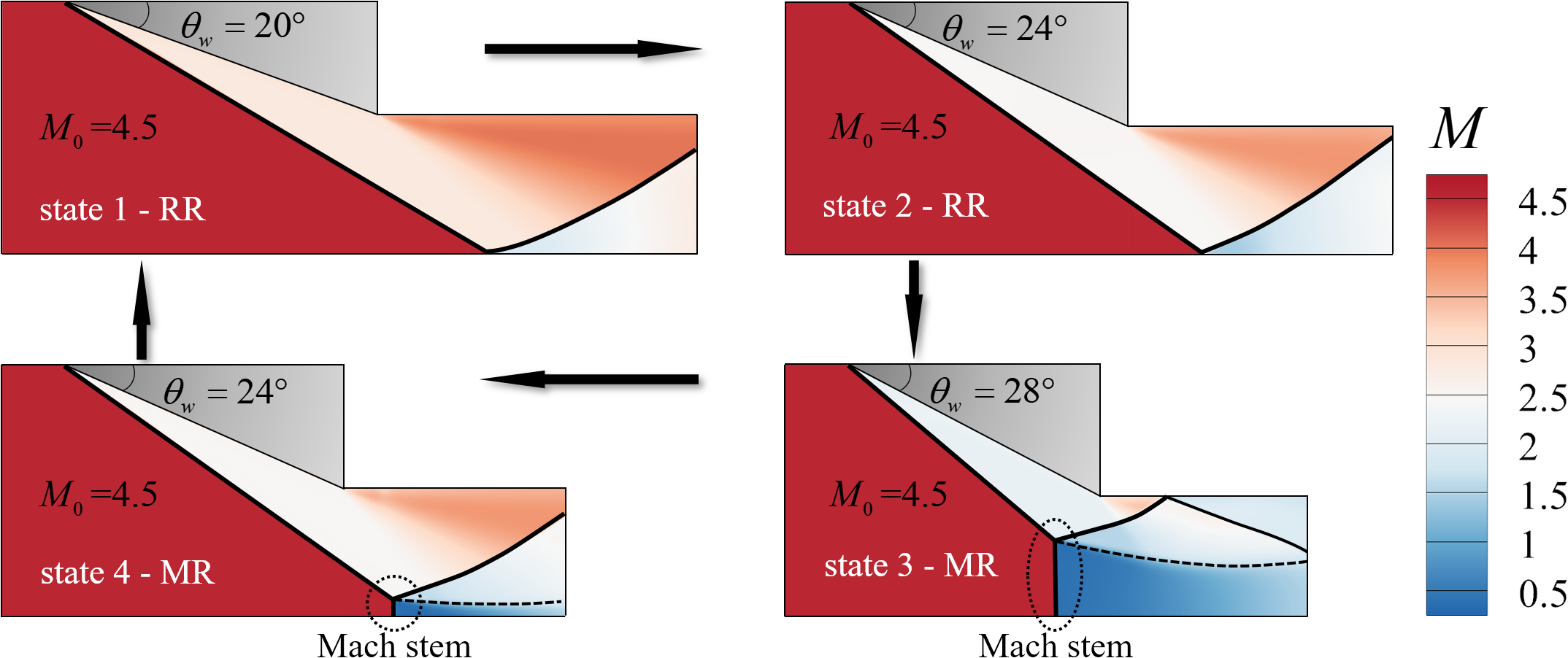}}
	\subfigure[\label{subfig:theta_boundary}]{\includegraphics[width = 0.49\columnwidth]{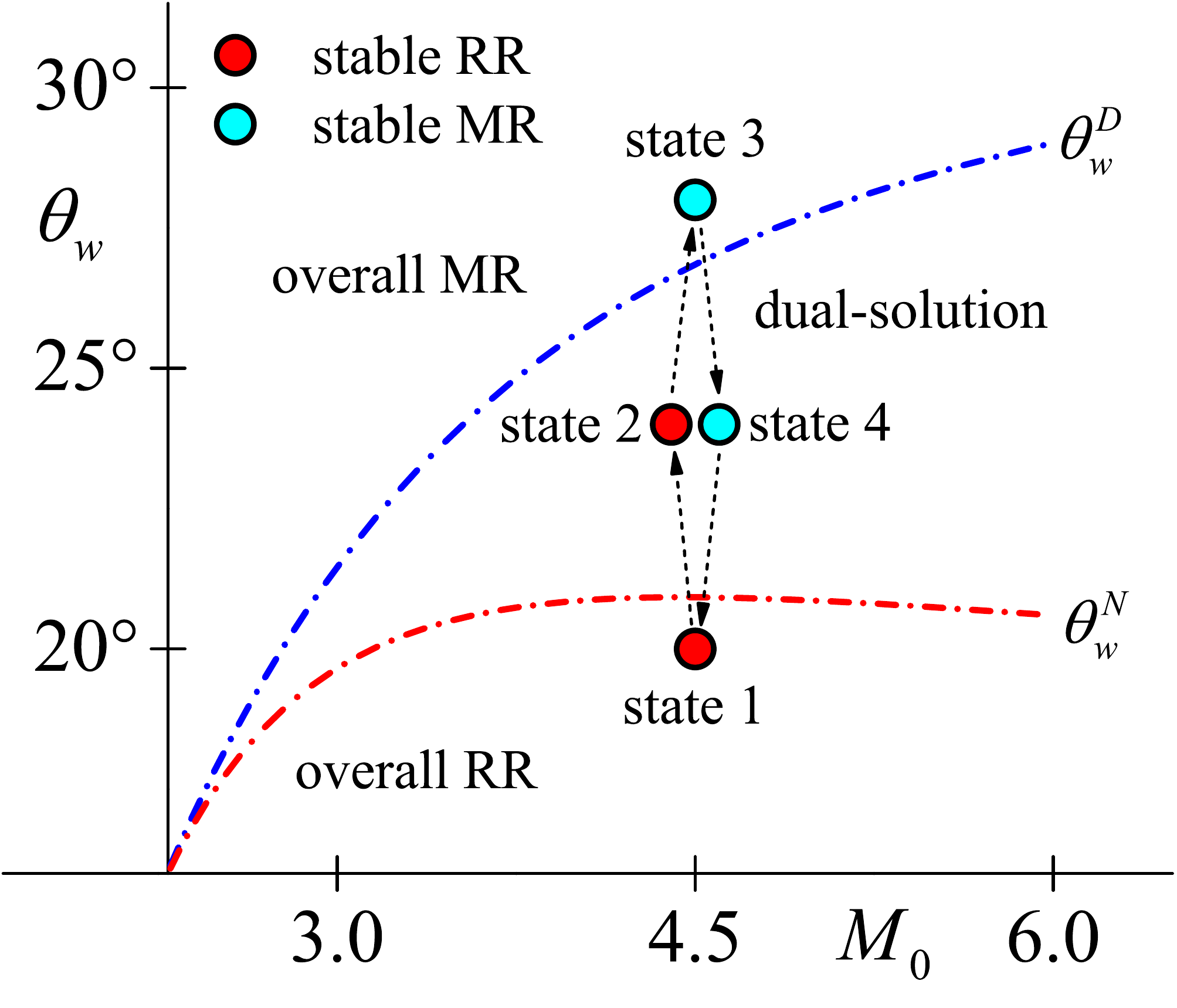}} 
	\subfigure[\label{subfig:shock_polar}]{\includegraphics[width = 0.49\columnwidth]{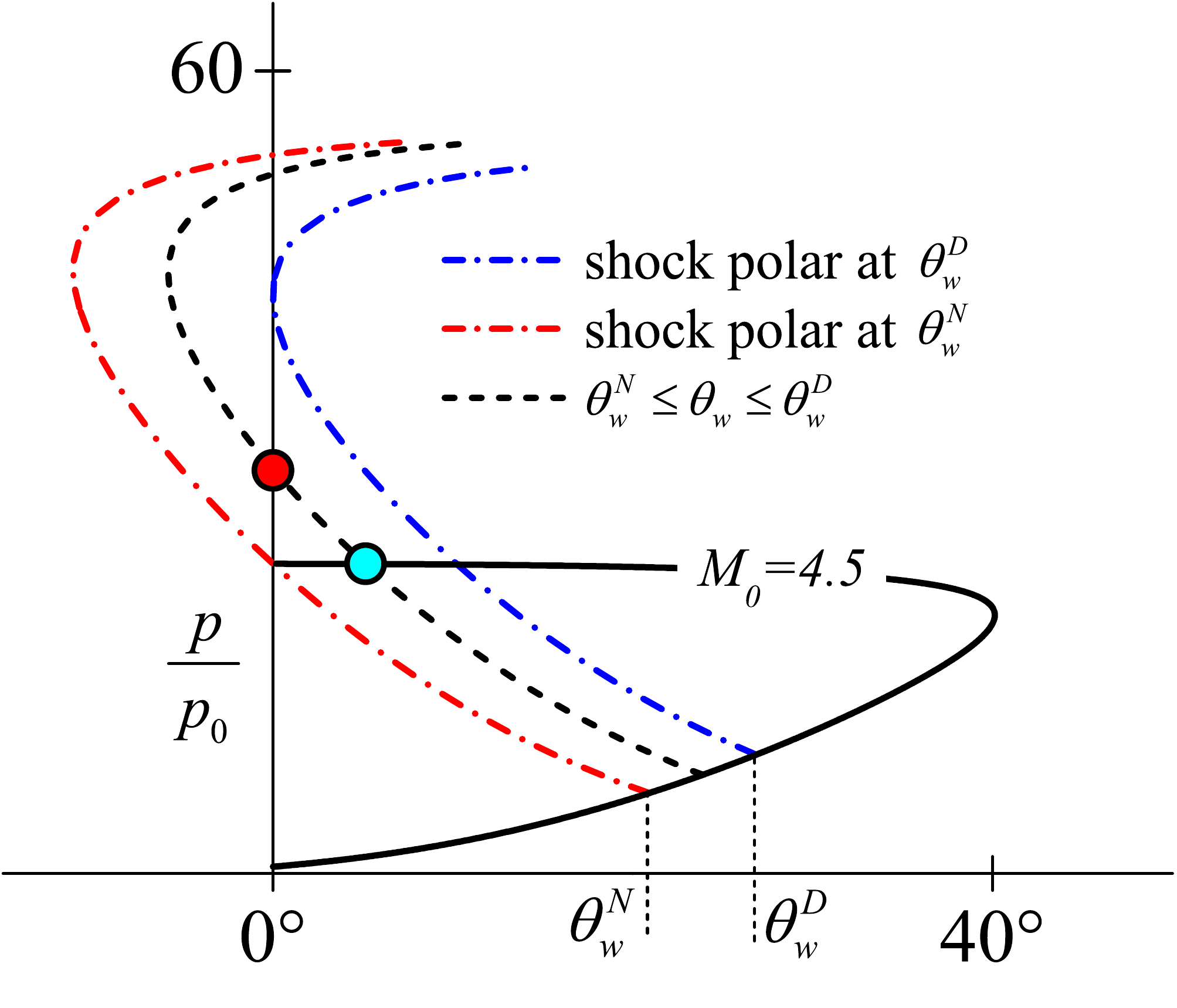}}
	\caption{(a) distributions of local Mach number $M$ of state 1,2,3 and 4 obtained by numerical computation at $M_{0} = 4.5$; (b) $\theta_w^D$ and $\theta_w^N$ varying with the inflow Mach number $M_{0}$, and the hysteresis loop (state 1, 2, 3 and 4) induced by $\theta_{w}$ variation at $M_{0}=4.5$; (c) shock polars at $\theta_w^D$, $\theta_w^N$ and $\theta_w$ at $M_{0}=4.5$, where $\theta_w$ is in the dual-solution domain. \label{fig:for_introduction}}
\end{figure}

 In shock wave reflections, which are ubiquitous in aerospace engineering, hystereses also exist in the transition between regular reflection (RR) and Mach reflection (MR), which two different configurations were first observed by Mach \cite{mach1878uber} in 1878, as shown in figure \ref{subfig:schlieren}, where RR corresponds to state 1, 2 and MR corresponds to state 3, 4, each of which has a segment of normal shock waves named Mach stem. More than half a century after that, von Neumann \cite{von1943oblique} proposed two critical deflection angles, the detachment condition $\theta_w^D$ and von Neumann condition $\theta_w^N$, for the RR$\rightleftharpoons$MR transition. As shown in figure \ref{subfig:theta_boundary} and \ref{subfig:shock_polar}, only RRs exist if the wedge angle $\theta_w$$<$$\theta_w^N$ and only MRs exist when $\theta_w$$>$$\theta_w^D$. However, in the intermediate range $\theta_w^N$$\leqslant$$\theta_w$$\leqslant$$\theta_w^D$, both stable RRs and MRs are theoretically possible, which makes range-$[\theta_{w}^{N},\theta_{w}^{D}]$ referred to as a dual-solution domain. Based on this fact, Hornung \cite{hornung_oertel_sandeman_1979} hypothesized that hystereses exist in RR$\rightleftharpoons$MR transitions, which was verified both experimentally \cite{chpoun1995reconsideration} and numerically \cite{vuillon1995reconsideration} later on, and re-initiated the researchers interest in the RR$\rightleftharpoons$MR transition, particularly, the hysteresis process \cite{hornung1982transition,ivanov1995hysteresis,ben2007shock}. Figure \ref{subfig:schlieren} shows flow configurations of hysteresis induced by $\theta_w$ variation corresponding to the states in figure \ref{subfig:theta_boundary}. As $\theta_w$ varies continuously from $20\degree$ (in the overall RR domain) to $24\degree$ (in the dual-solution domain), the configuration maintains stable RR (from state 1 to 2). However, if $\theta_w$ varies from $28\degree$ (in the overall MR domain) back to $24\degree$, it will maintain stable MR (from state 3 to 4). It is noted that at $\theta_w = 24\degree$, either stable RR or MR appearing depends on the evolution history. Obviously, state 1 $\rightarrow$ 2 $\rightarrow$ 3 $\rightarrow$ 4 $\rightarrow$ 1 constitutes a hysteresis loop. Although forty years has passed sicne the hysteresis was put forward \cite{hornung_oertel_sandeman_1979}, the mechanism behind this complex phenomenon is still not clear.
\par In this letter, the least-action principle is used to reveal the essence of hysteresis in the RR$\rightleftharpoons$MR transition, i.e. how the reflection depends on its evolution history and where the transition happens. 
\par
\begin{figure}
	\subfigure[\label{subfig:illustration_MR}]{\includegraphics[width = 1.0\columnwidth]{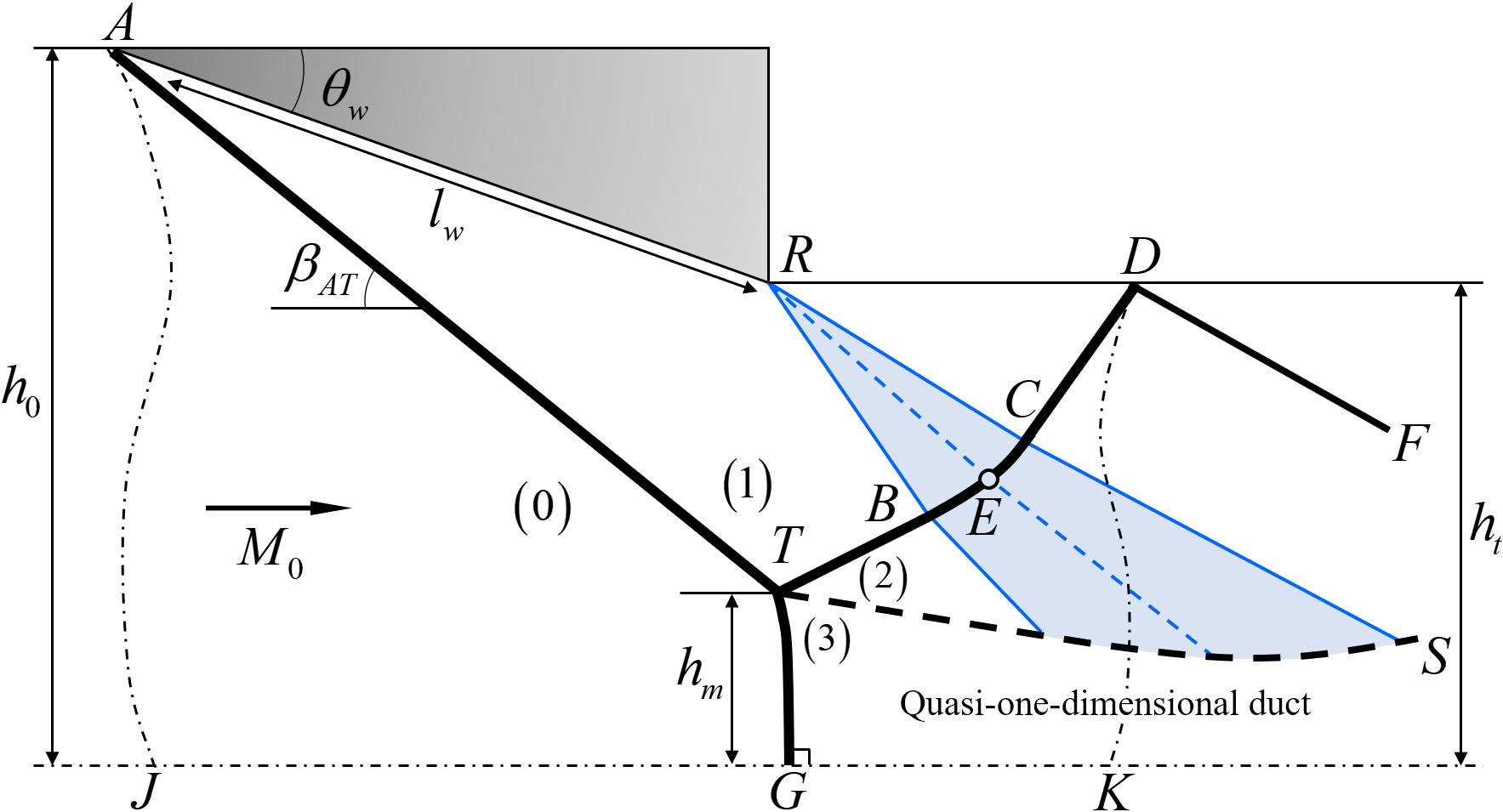}}
	\subfigure[\label{subfig:beta_RR-MR}]{\includegraphics[width = .49\columnwidth]{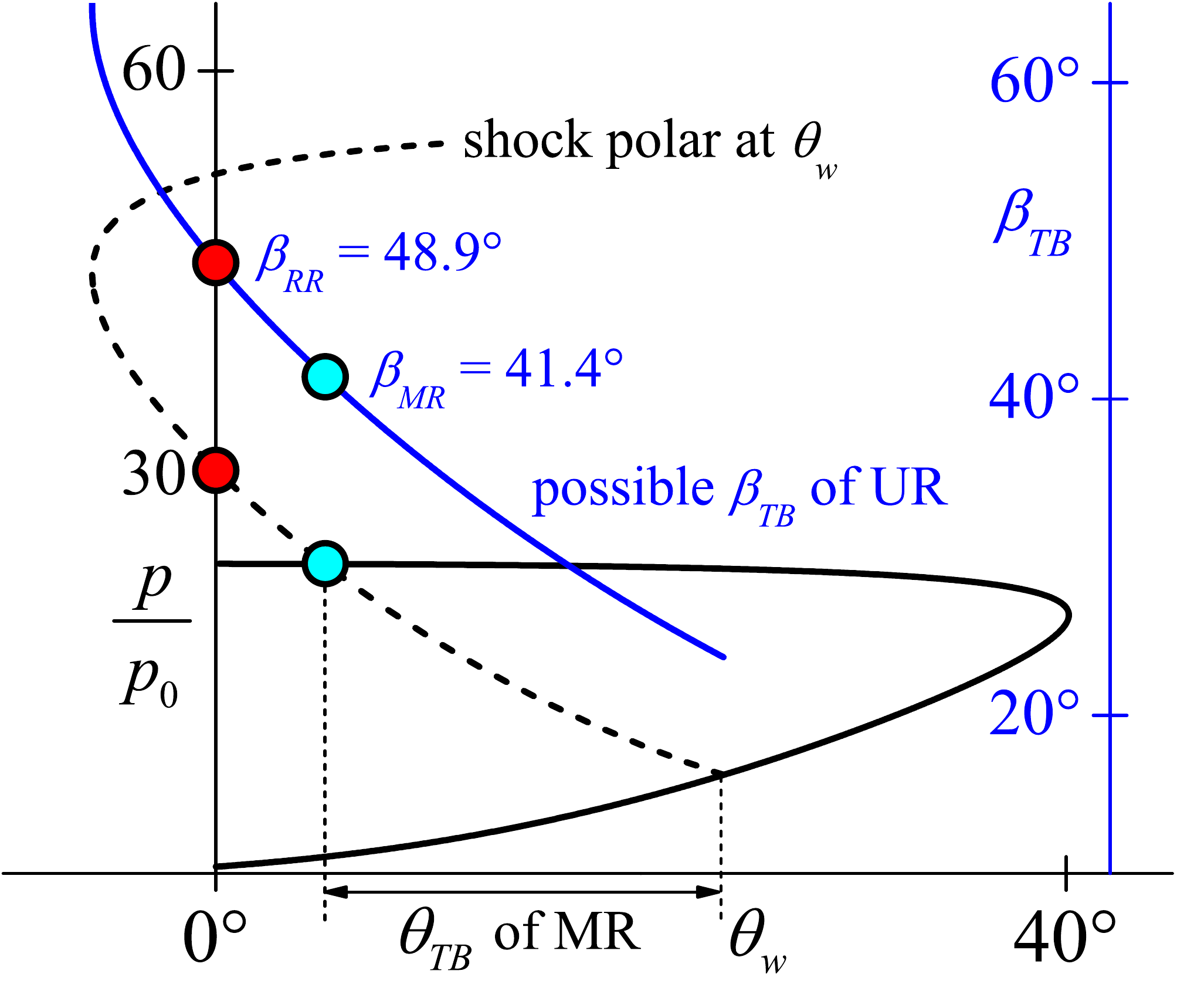}}
	\subfigure[\label{subfig:beta_Mach_height}]{\includegraphics[width = .49\columnwidth]{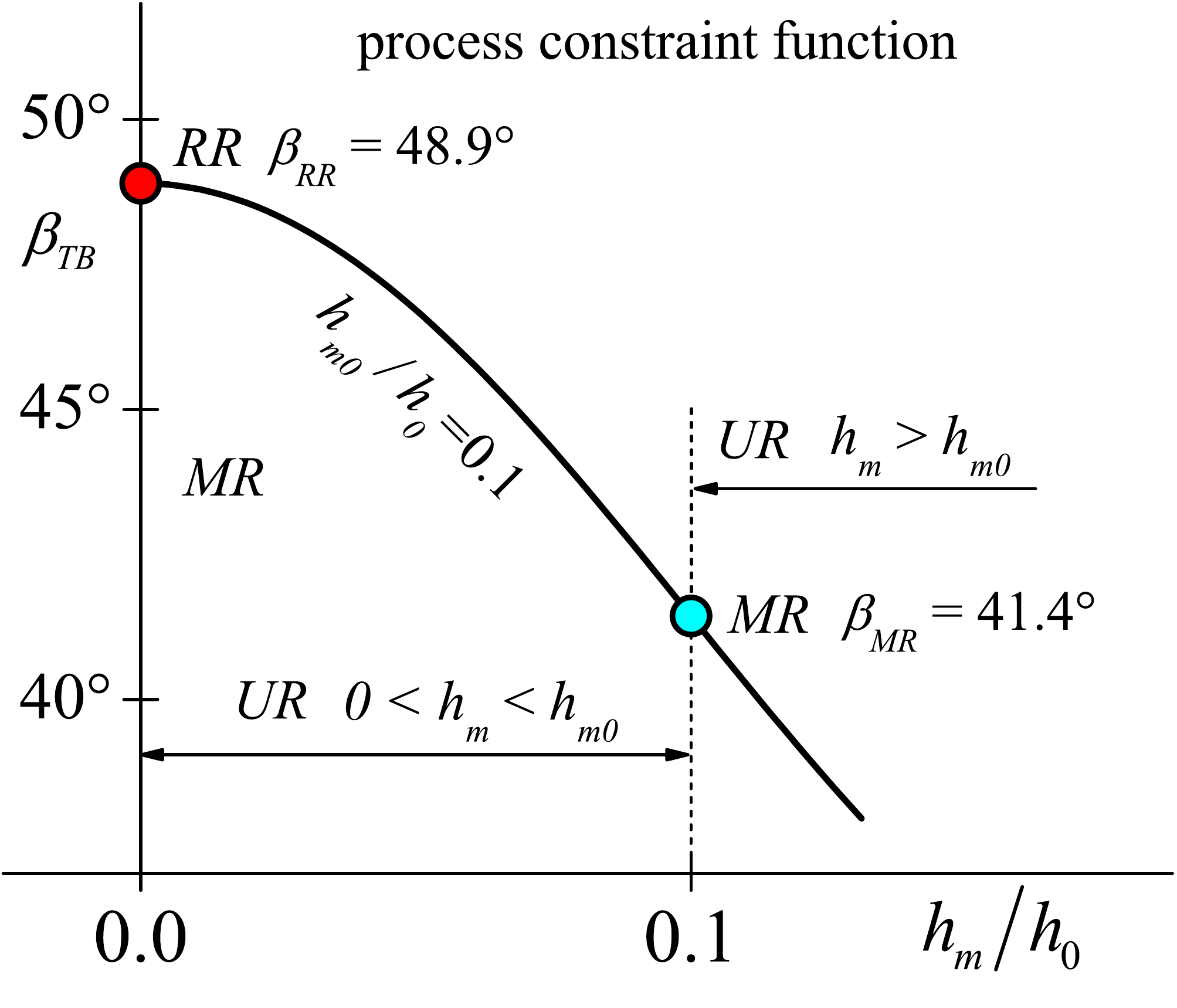}}
	\caption{(a) the illustration of a general reflection configuration with a Mach stem height $h_{m}$; (b) possible $\beta_{TB}$ of unstable reflection configurations; (c) process constraint function of $\beta_{AT}$ with the order parameter $h_m$. \label{fig:illustration_beta_RR-MR_beta_Mach_height}}
\end{figure}

 First, we will introduce the reflection configuration. As shown in figure \ref{subfig:illustration_MR}, for a given wedge angle $\theta_{w}$ in the dual-sultion domain, a general reflection configuration is composed of (i) an incident shock wave $AT$, (ii) a strong shock wave called the Mach stem $TG$ with height $h_m$ ($h_m = 0$ corresponds to a RR and $h_m > 0$ corresponds to a MR) and (iii) a reflected shock wave including a straight segment $TB$, a curved segment $BC$ and another straight segment $CD$. It is notice that, in analogy with the magnetization $\kappa$ manifesting the ferromagnet state in the ferromagnetism theory \cite{MichaelPlischke2006Equilibrium,goldenfeld2018lectures}, the Mach stem height $h_m$ also manifests the flow state in the shock wave reflection, then it is actually the order parameter of this system, and $\theta_{w}$ is the external parameter analogous with the magnetic field intensity $H$. For a MR ($h_{m} > 0$), a free shear layer $TS$ exists, below which the flow can be regarded as Quasi-one-dimensional isentropic duct flow. Behind the reflected shock wave $TD$, the flow is expanded or weakly compressed, which can also be regarded as isentropic \cite{gao2010study,bai2017size}.
\par Then we will demonstrate that the reflection flow system has the minimal dissipation. The Helmholtz-Rayleigh dissipation theorem \cite{helmholtz1868theorie,rayleigh1913lxv,serrin1959mathematical} put forward that an incompressible viscous fluid should have the minimal dissipation if the acceleration $\mathbf{a}=\mathbf{u} \cdot \nabla \mathbf{u}$ can be derived by a potential $\zeta$, i.e. $\mathbf{a}=\nabla \zeta$ or $\nabla \times \mathbf{a} = 0$. This theorem was extended to compressible flows by He et al \cite{ho1988extension,wu2007vorticity} in 1988. Mathematically, it means that the steady compressible Navier-Stokes equation:
\begin{equation}
\mathbf{u} \cdot \nabla \mathbf{u}=-\frac{1}{\rho}\nabla p +\mathbf{f}+\frac{1}{\rho}{[\nabla(\eta \vartheta)+\nabla \cdot(2 \mu \mathbf{D})] \label{eq:NS-equation}}
\end{equation}
can be derived by the variation of the dissipation. In function (\ref{eq:NS-equation}), $\mathbf{u}$, $p$, $\rho$ , $\mathbf{f}$, $\eta$ and $\mu$ are the velocity, pressure, density, body force, dilatation and shear viscosity of the flow, respectively. $\vartheta=\nabla \cdot \mathbf{u}$ and $\mathbf{D}=\left[\nabla \mathbf{u}+(\nabla \mathbf{u})^{T}\right] / 2$ are the divergence and the strain-rate tensor, respectively. $\phi=\eta \vartheta^{2}+2 \mu \mathbf{D} : \mathbf{D}$  is the kinetic energy dissipation \cite{wu2007vorticity}. We consider the total dissipation $\Phi$ in control volume $V$ bounded by $\ell$, which satisfies that $V$ is nondeformable or the flow on mobile $\ell$ (if $V$ is deformable) is nondissipative. With the constraint served by the steady continuity equation $\nabla \cdot(\rho \mathbf{u})=0$, the variation of $\Phi$ can be written as:
\begin{equation}
\delta \Phi=\delta \int_{V}[\phi+\lambda \nabla \cdot(\rho \mathbf{u})] d V=0
\end{equation}
where $\lambda$ is Lagrangian multiplier, and $L=\phi+\lambda \nabla \cdot(\rho \mathbf{u})$ is the Lagrangian. Since $\mathbf{u}$ and $\rho$ are the two independent variables of $L$, the Eular-Lagrangian equation is:
\begin{align}
& \frac{\delta L}{\delta \mathbf{u}}=0 :\quad[\nabla(\eta \vartheta)+\nabla \cdot(2 \mu \mathbf{D})]+\frac{1}{2} \rho \nabla \lambda=0  \label{eq: varation u}  \\ 
& \frac{\delta L}{\delta \rho}=0 :\quad \mathbf{u} \cdot \nabla \lambda=0 \label{eq: varation rho} 
\end{align}
If a flow satisfies that (i) $\mathbf{a}=\nabla \zeta$; (ii) $\mathbf{f} = \nabla U$, i.e. the body force can be derived by a potential $U$; (iii) $\nabla p / \rho = \nabla \int d p / \rho$, i.e. the flow is barotropic which is equivalent to $\nabla p \times \nabla \rho=0$, then when $\lambda$ is chosen as $\lambda=-2\left(\zeta+\int d p / \rho + U\right)$, formular (\ref{eq: varation u}) can be exactly rearranged to function (\ref{eq:NS-equation}), and formular (\ref{eq: varation rho}) is the Bernoulli Integration. Therefore, a flow satisfying (i), (ii) and (iii) has the minimal dissipation.
\par For a flow passing through a straight shock wave, the acceleration $\mathbf{a}$ can be decomposed into two parts relative to the shock front, i.e. the normal component $a_{n}$ and tangential one $a_{\tau}$ (consider a 2D flow field). Since the velocity only changes perpendicularly through the shock, there must be $\partial a_{n} / \partial \tau = 0$ and $a_{\tau} = 0$, then $\left|\nabla \times \mathbf{a}\right| = \partial a_{\tau} / \partial n - \partial a_{n} / \partial \tau = 0$, which implies that condition (i) is satisfied. Additionally, the body force $\mathbf{f}$ is gravity that can be negligible, which means condition (ii) is satisfied. $\nabla p$ and $\nabla \rho$ are both perpendicular to the shock front, which implies $\nabla p \times \nabla \rho=0$ and then condition (iii) is satisfied. Therefore, a steady flow across a straight shock wave has minimal dissipation. This is also the reason that, although both a weak and a strong oblique shock wave are theoretically possible at a same deflection angle, the observable shock wave in reality is always the weak one. Furthermore, if a shock wave is curved, its curvature radius is always infinitely great compared to its thickness, then it can also be approximated to a stright one.  Therefore, if the total dissipation of a steady flow field is only contributed by shock waves, this flow should have minimal dissipation.
\par For the flow system shown in figure \ref{subfig:illustration_MR}, a control volume $V$ is chosen to enclose the reflection configuration, of which boundaries $\ell$ are composed of the upper wall $ARD$, the reflected surface $JK$, the inflow surface $AJ$ and the outflow surface $DK$. Except shock waves, dissipation could also happen in (i) the isentropic region and (ii) the shear layer $TS$. For the isentropic region, the relationship of $\phi$ and entropy generation $d s / d t$ is that $\phi=\mathcal{T} d s / d t$ \cite{wu2007vorticity}, where $\mathcal{T}$ is the flow temperature, then "isentropic" $d s / d t = 0$ means "nondissipative" $\phi = 0$. For the shear layer, the dissipation induced by (ii) $TS$ and a shock wave are $\int_{\varepsilon_{\delta}} \phi d x \sim \mu \Delta u^{2} / \varepsilon_{\delta}$ and $\int_{\varepsilon} \phi d x \sim \mu \Delta u^{2} / \varepsilon$, respectively, where $\varepsilon_{\delta}$ and $\varepsilon$ are the thickness of $TS$ and the shock wave, respectively, and $dx$ is the normal infinitesimal length of $\varepsilon_{\delta}$ and $\varepsilon$. Since $\varepsilon_{\delta}$ is always much larger than $\varepsilon$ of a shock wave, i.e. $\varepsilon_{\delta} \gg \varepsilon$, there must be $\mu \frac{\Delta u^{2}}{\varepsilon_{\delta}} \ll \mu \frac{\Delta u^{2}}{\varepsilon}$, i.e. the dissipation induced by $TS$ is negligible. Additionally, the flow on the mobile boundary $DK$ is nondissipative. Therefore, $\Phi$ in $V$ is mainly contributed by the five shock waves:
\begin{equation}
\Phi = \Phi_{AT} + \Phi_{TB} + \Phi_{BC} + \Phi_{CD} + \Phi_{TG}
\label{eq: total_dissipation_function}
\end{equation}
where $\Phi_{AT}$, $\Phi_{TB}$, $\Phi_{BC}$, $\Phi_{CD}$ and $\Phi_{TG}$ are the dissipation induced by $AT$, $TB$, $BC$, $CD$ and $TG$, respectively. Thus, the synergy \cite{haken1977synergetics} of these shock waves should make the flow system have the minimal dissipation. 
\par Next we will expound that $\Phi$ only depends on the order parameter $h_{m}$ for a given external parameter $\theta_{w}$. Since the viscous dissipation is the dominate term of the kinetic energy loss in compressible flows \cite{guarini2000direct,pirozzoli2004direct}, the dissipation $\phi$ of a oblique shock wave per unit length can be approximate to:
\begin{equation}
\phi \simeq \frac{1}{2}\left[\rho_{a}\left(M_{a} \mathcal{A}_{a} \sin \beta\right)^{3} - \rho_{b}\left[M_{b} \mathcal{A}_{b} \sin \left(\beta-\theta\right)\right]^{3}\right]
\label{eq: dissipation_function}  
\end{equation}
where subscript ‘$a$’ and ‘$b$’ denote variables ahead of and behind the shock wave, respectively. $\theta$ and $\mathcal{A}$ are the flow defection angle across the shock wave and the acoustic velocity, respectively. $\beta$ is the shock angle that satisfies:
\begin{equation}
f_{\beta}(M_{a}, \beta, \theta)=2 \cot \beta \frac{M_{a}^{2} \sin ^{2} \beta-1}{M_{a}^{2}(\gamma+\cos 2 \beta)+2}-\tan \theta = 0
\label{eq: Ma_beta_theta}
\end{equation}
which means that, for a given Mach number $M_{a}$, either of $\beta$ or $\theta$ is known, the other one can be obtained. Furthermore, variables on both sides of a shock front satisfy the classical Rankine-Hugoniot (RH) relations \cite{rankine1870xv,rankine1887propagation,salas2007curious}:
\begin{equation}
\Omega_{b}=\Omega_{a} * \mathrm{X}\left(M_{a}, \beta\right)
\label{eq: R_H_relation}
\end{equation}
where $\Omega = [M^{2}, \rho, \mathcal{A}]^T$, $\mathrm{X} = [f_{M}, f_{\rho}, f_{\mathcal{A}}]^T$ are namely the RH ralations, and `$*$' denotes hadamard product. Therefore, $\Omega_{b}$ and $\phi$ can be obtained for given $\Omega_{a}$ and $\theta$ with (\ref{eq: dissipation_function} - \ref{eq: R_H_relation}). For the straight incident shock wave $AT$, $\Phi_{AT} = \phi_{AT}\sigma_{AT}$, where $\sigma_{AT}$ is the length of $AT$ and determined by $h_{0}$, $h_{m}$ and $\beta_{AT}$. $\phi_{AT}$ can be obtained with $\theta_{w}$ and $\Omega_{AT,a} = [M_{0}^{2}, \rho_{0}, \mathcal{A}_{0}]^T$. For the Mach stem $TG$, $\Phi_{TG} = \int_{T}^{G} \phi d \sigma$. Since $TG$ is just slightly bend, it can be approximate to a straight one, i.e. $\Phi_{TG} \simeq \frac{1}{2}\left(\phi_{T(3)} + \phi_{G}\right)h_{m} $, where subscript ‘$(3)$’ refers to the location near the triple point $T$ in figure \ref{subfig:illustration_MR}. As $\theta_{G} = 0$, $\Omega_{a,G} = \Omega_{a,T(3)} = \Omega_{a,AT}$ and $\theta_{T(3)}$ can be calculated by the three-shock theory \cite{von1945refraction}. Although the Mach stem is moving, its velocity is small compared to the inflow \cite{Mouton2007Mach}. Therefore, unstable reflections are approximate to quasi steady, which implies (\ref{eq: dissipation_function} - \ref{eq: R_H_relation}) is still available, thus $\phi_{G}$ and $\phi_{T(3)}$ can be obtained. For the reflection shock wave $TB$, $\Phi_{TB} = \phi_{TB}\sigma_{TB}$. stable RR ($h_{m} = 0$) and stable MR ($h_{m} = h_{m0} > 0$) correspond to $\beta_{TB} = \beta_{RR}$ and $\beta_{TB} = \beta_{MR}$ , respectively, where $\beta_{RR}$ and $\beta_{MR}$ can be obtained by the two- and three-shock theory \cite{von1943oblique,von1945refraction}, which are shown in figure \ref{subfig:beta_RR-MR}. Supposing a transition process of an unstable reflection (UR) from the stable RR to stable MR, $\beta_{TB}$ should monotonously decrease from $\beta_{RR}$ to $\beta_{MR}$ while $h_{m}$ monotonously increase from 0 to $h_{m0}$. Considering the physical reality that the shear layer $TS$ should not touch the reflected surface $JK$ \cite{Mouton2007Mach} when the stable RR just changes into an UR, $\beta_{TB}$ must maintain $\beta_{RR}$ when $h_{m} = 0 + d h_{m}$, i.e. $(d \beta_{T B} / d h_{m})_{0}=0$. When the UR is near the stable MR with $h_{m} = h_{m0} \pm d h_{m}$, $\beta_{T B}$ is assumed varying linearly, i.e. $(d \beta_{T B} / d h_{m})_{h_{m0}} = $ constant. Therefore, $\beta_{TB}$ can be approximate to:
\begin{equation}
\frac{\beta_{T B}-\beta_{M R}}{\beta_{R R}-\beta_{M R}}=f_{\beta_{TB}} (h_{m})=\cos \left(\frac{h_{m}}{h_{m 0}} \cdot \frac{\pi}{2}\right)
\end{equation}
where function $f_{\beta_{TB}}$ is a phenomenological relation to constrain the variation process of $\beta_{TB}$ with the order parameter $h_{m}$. As $\beta_{TB}$ can describe the possible unstable state of the system, we call $\beta_{TB}$ and $f_{\beta_{TB}}$ as `the process parameter' and `the process constraint function', respectively, which are shown in figure \ref{subfig:beta_Mach_height}. Then $\phi_{TB}$ depending on $h_{m}$ (originally depending on $\beta_{TB}$) can be obtained with $\Omega_{TB,a} = \Omega_{AT,b}$. As $B$ is the intersection of $TB$ and the Mach wave $RB$, $\sigma_{TB}$ also depends on $h_{m}$ with geometrical relations. For the shock wave $BC$, which is curved by the Prandtl–Meyer expansion fan \cite{courant1999supersonic,von2004mathematical}, $\Phi_{BC} = \int_{B}^{C} \phi_{E} d \sigma$, where $E$ is the point moving from point $B$ to $C$. Bai et al \cite{bai2017size} has derived the differential relations of $\beta_{E}$ and $(x_{E},y_{E})$, with which they obtained $\Omega_{E,a}$ and the shape of $BC$. Based on their theory, we can obtain $\Phi_{BC}$. For shock wave $CD$, $\Phi_{CD} = \phi_{CD}\sigma_{CD}$. Since $\beta_{CD}=\beta_{CE}$ and $\Omega_{CD,a}=\Omega_{C,a}$, we can obtain both $\phi_{CD}$ and $\sigma_{CD}$. Thus, the total dissipation $\Phi_{CD}(\theta_{w},h_{m})$ induced by $CD$ can be obtained.
\begin{figure*}
	\centering
	\subfigure[\label{subfig:3d_line}]{\includegraphics[width = 0.5\columnwidth]{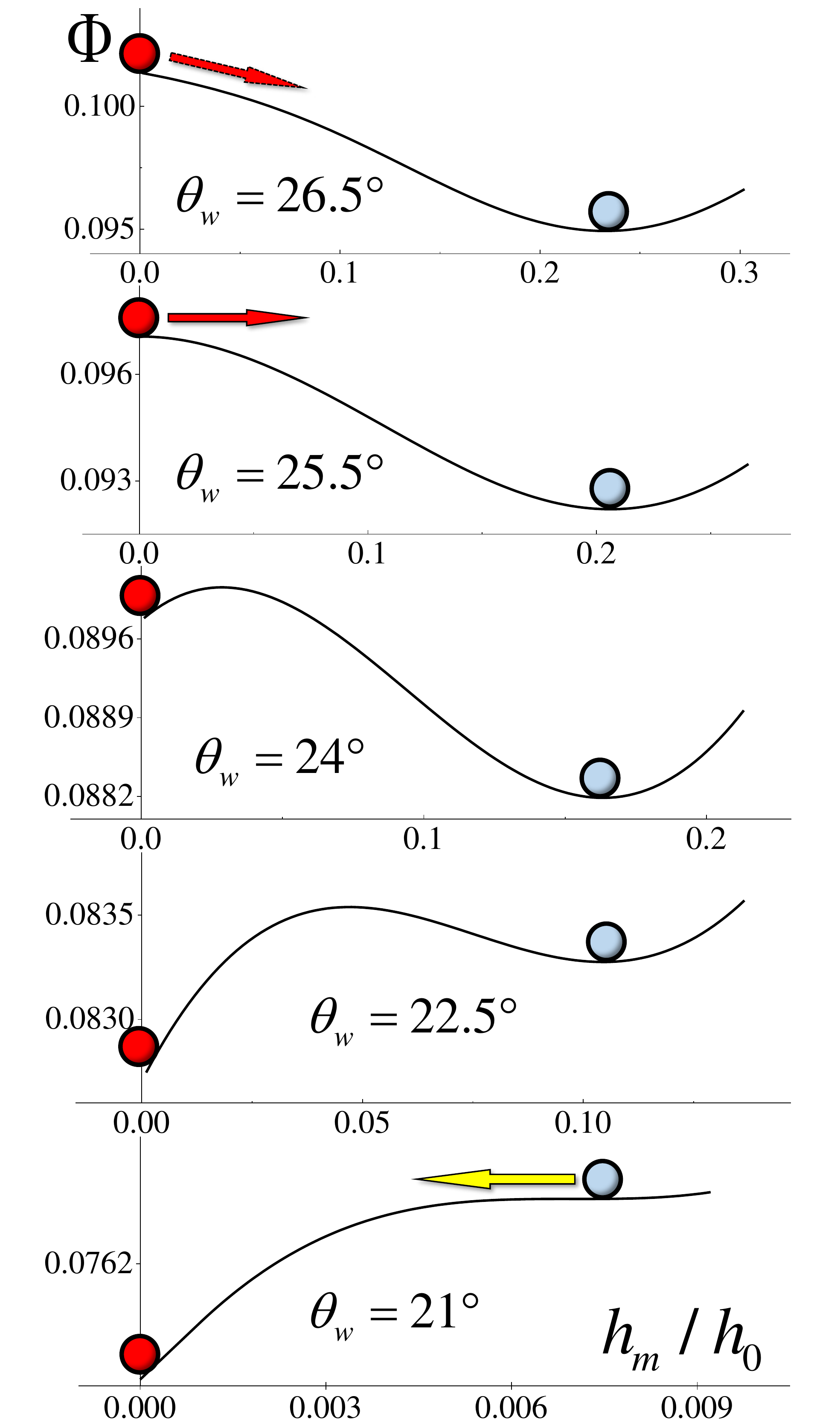}}
	\subfigure[\label{subfig:3d_surface}]{\includegraphics[width = 0.825\columnwidth]{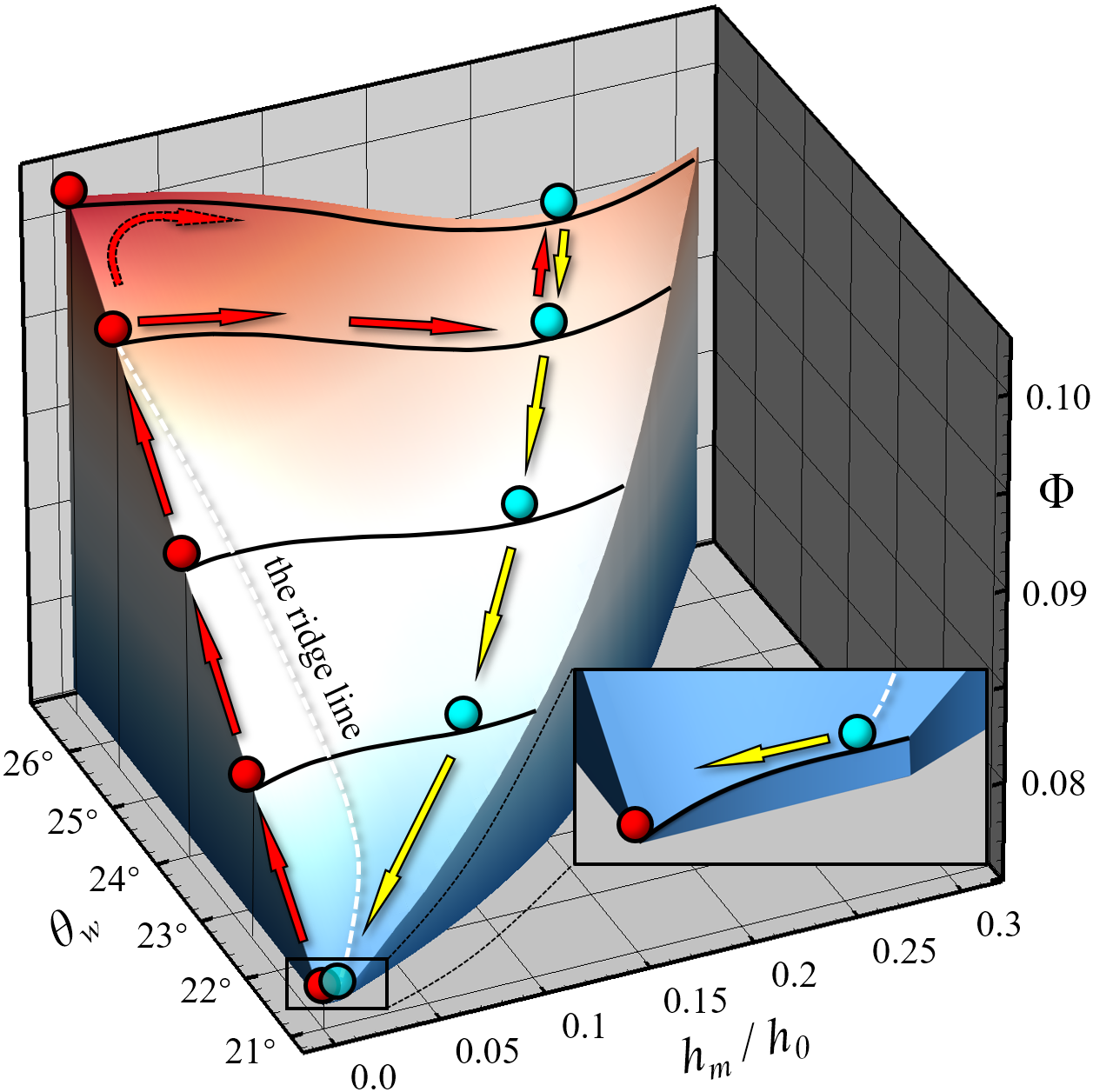}}
	\subfigure[\label{subfig:3d_2dsurface}]{\includegraphics[width = 0.575\columnwidth]{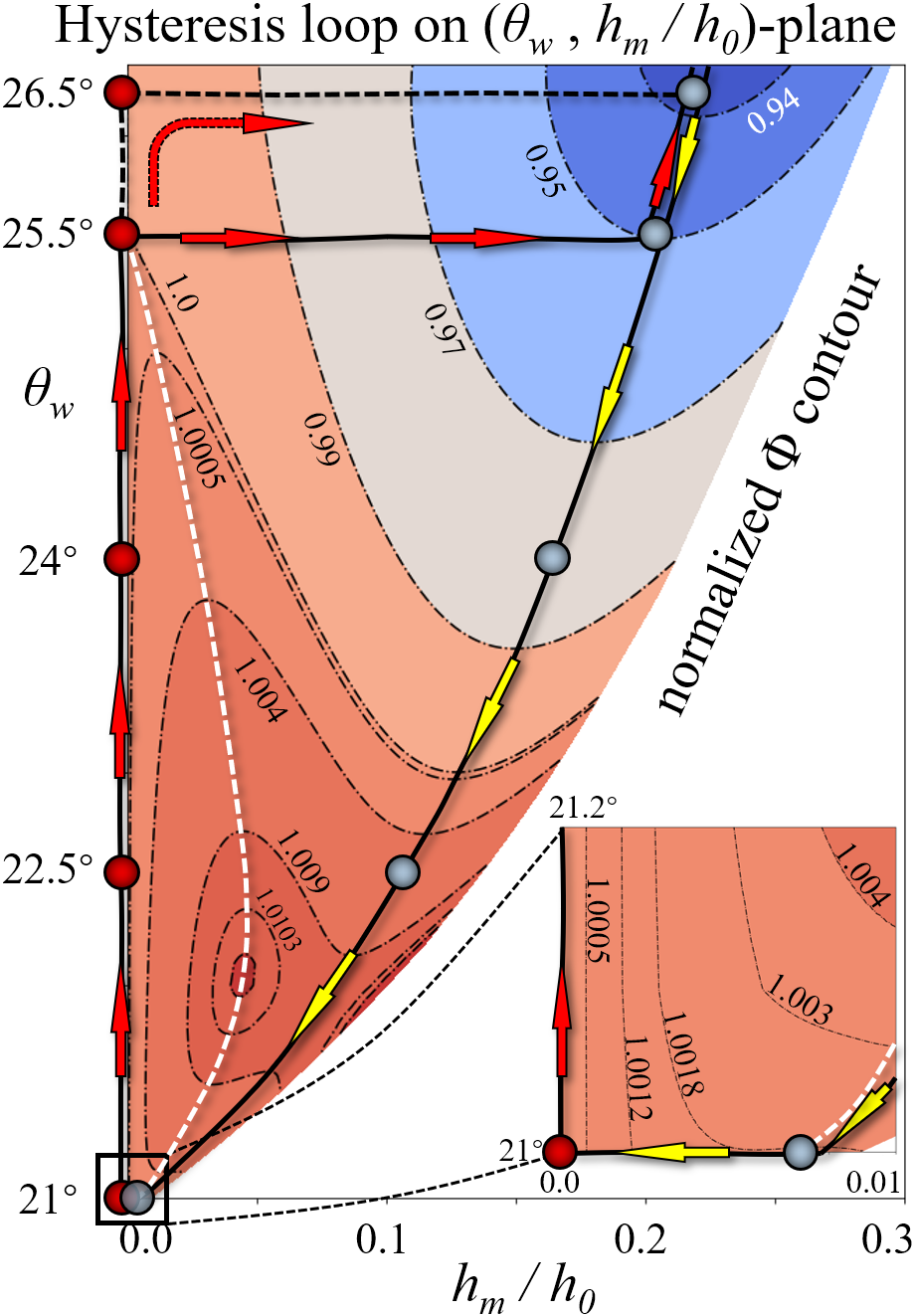}}
	\caption{different perspectives of the total dissipation $\Phi$ landscape, where red and blue spheres correspond to stable RRs and MRs, respectively. (a) five latitude lines from the front view of $\Phi$ landscape, where $\Phi$ versus normalized order parameter $h_{m} / h_{0}$ at $\theta_w = $ 21$\degree$, 22.5$\degree$, 24$\degree$, 25.5$\degree$ and 26.5$\degree$; (b) the $\Phi$ landscape, where red and yellow arrows are the path of RR $\rightarrow$ MR and RR $\leftarrow$ MR in the valley, respectively, and the white dush line is the ridge line; (c) contour of normalized total dissipation $\Phi / \Phi(\theta_{w}, 0)$, where the black solid line is the hysteresis loop of the RR$\rightleftharpoons$MR transition, and the black dash line is the path when $\theta_{w} > 25.5\degree$. \label{fig:3d_dissipation_surface}}
\end{figure*}
\par Now we have the total dissipation $\Phi$ with formular (\ref{eq: total_dissipation_function}). The synergy principle \cite{haken1977synergetics} of shock waves constituting a stable configuration, is that they must make the flow maintain the minimal dissipation. Thus, for a given $M_0$, a stabel Mach stem $h_{m0}$ should satisfy:
\begin{equation}
\frac{\partial \Phi}{\partial h_m}\bigg|_{h_{m0}}=0,\quad \frac{\partial^2 \Phi}{\partial h_m^2}\bigg|_{h_{m0}}>0 
\label{eq: local_min_differential_relation}
\end{equation}
\par The landscape of $\Phi(\theta_w,h_{m} / h_{0})$ is shown in figure \ref{subfig:3d_surface} with $\theta_w$ varying in the dual-solution domain at $M_0=4.5$. As shown in figure \ref{subfig:3d_line}, the initial state is the stable RR corresponding to $\Phi(21\degree,0)$, where the possible MR is highly unstable but the RR is stable, which means that a disturbance can not transform the RR to the MR, but can easily transform the MR to the RR. As $\theta_{w}$ increases to 22.5$\degree$, a dissipation barrier emerge and two minimal values will be formed corresponding to two stable configurations, i.e. the stable RR and MR at $\Phi(22.5\degree,0)$ and $\Phi(22.5\degree,h_{m0})$, respectively. If $\theta_w$ varies continuously and slowly, the disturbance will be not strong enough to transform the RR to the MR, then the configuration maintains the stable RR. It is similar as $\theta_{w}$ increases to 24$\degree$, the only difference is that $\Phi(24\degree,0) > \Phi(24\degree,h_{m0})$, but $\Phi(22.5\degree,0) < \Phi(22.5\degree,h_{m0})$. When $\theta_{w}$ increases to 25.5$\degree$, the RR becomes unstable, and just a little disturbance can transform it to the MR. Once $\theta_{w} > 25.5\degree$, the configuration will stay at the stable MR. As shown in figure \ref{subfig:3d_surface} and \ref{subfig:3d_2dsurface}, the transition point $\theta_{w} = 25.5\degree$ is a saddle-node bifuraction of the $\Phi$ landscape, i.e. the intersection point of the valley line $\Phi (\theta_{w} \le 25.5\degree, h_{m} = 0)$ and the ridge line. If $\theta_{w}$ decreases back from 26.5$\degree$,  although stable RRs are possible theoretically when $\theta_{w} < 25.5\degree$, the configuration will always stay at stable MRs until it reaches to 21$\degree$, where a little disturbance will transform the MR to the RR. The transition point $\theta_{w} = 21\degree$ is another saddle-node bifuraction, i.e. the intersection point of the valley line $\Phi (\theta_{w} \le 26.5\degree, h_{m} = h_{m0})$ and the ridge line. As $\theta_w$ varies from 21$\degree$ to 26.5$\degree$ and then back to 21$\degree$, a 3D path in the valleies of $\Phi(\theta_w,h_m)$ landscape emerges, which manifests a series of stable configurations. Obviously, the projection of the path to the ($\theta_w$,$h_m / h_{0}$)-plane is the 
hysteresis loop, as shown in figure \ref{subfig:3d_2dsurface}. In addition, it is seen that the MR$\rightarrow$RR transition occurs very close to $\theta_w^N=20.92\degree$, while the RR$\rightarrow$MR transition takes place at about 25.5$\degree$, which is smaller than $\theta_w^D=26.85\degree$. This phenomenon, i.e the transition point smaller than $\theta_w^D$, was also observed by Chpoun et al \cite{chpoun1995reconsideration}, which is clarified now that the MR has a larger stable region than the RR in the dual-solution domain.
\begin{figure}
	{\includegraphics[width = 0.95\columnwidth]{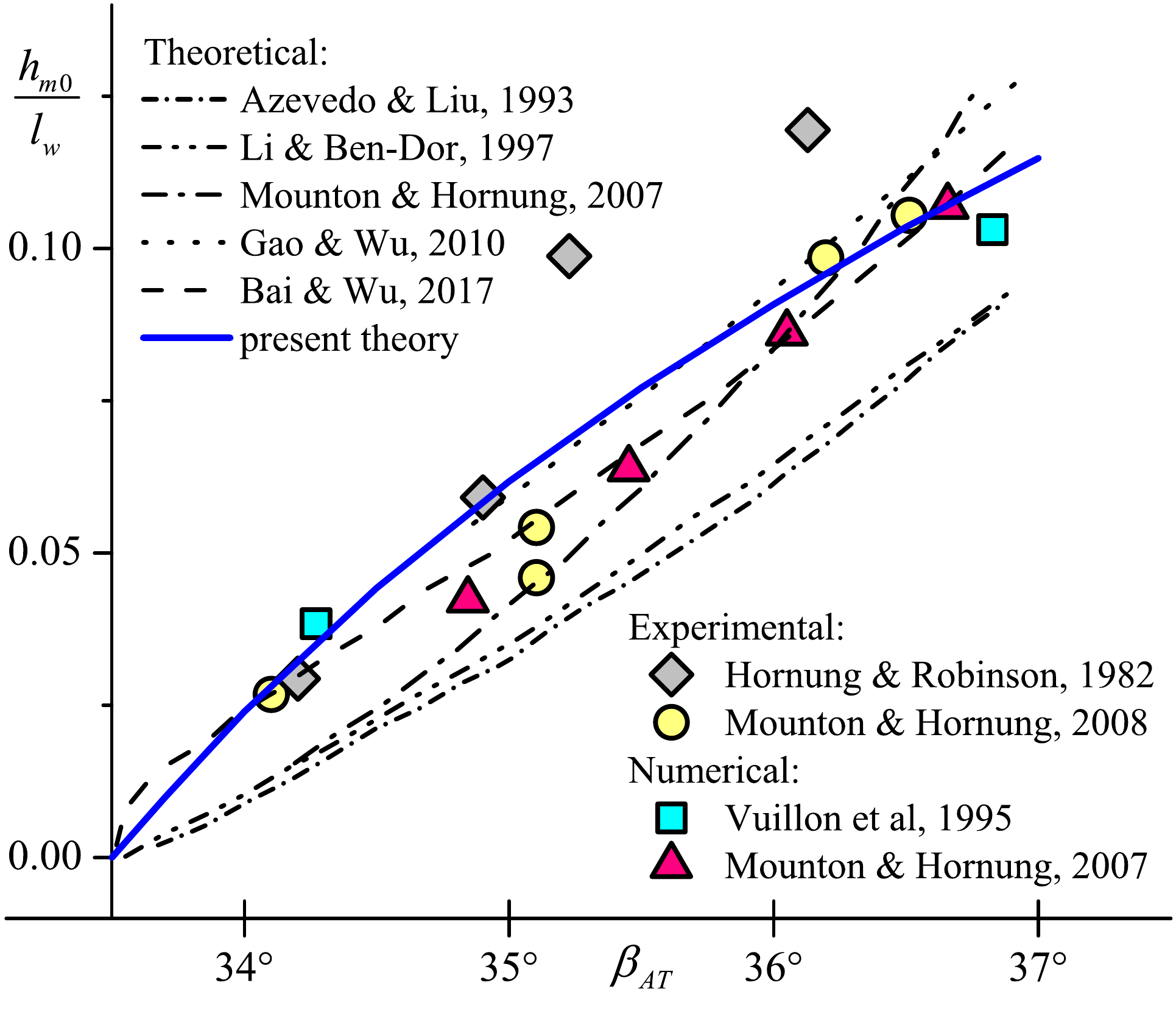}}
	\caption{comparison of the present theory with previous works at $M_0=3.98$ and $h_R/l_w \approx 0.4$, including the experimental results from Hornung et al \cite{hornung1982transition, Mouton2008Experiments}, the numerical results from Mouton \& Hornung \cite{Mouton2007Mach} and Vuillon et al \cite{vuillon1995reconsideration}, and the theoretical results from Azevedo \& Liu \cite{Azevedo1993Engineering}, Li \& Ben-Dor\cite{li1997parametric} ,Mouton \& Hornung \cite{Mouton2007Mach} and Wu et al \cite{gao2010study, bai2017size}. \label{fig:comparison}}
\end{figure}
\par Further on, as a validation, we compare $h_{m0}$ obtained by the present theory with previous experimental, numerical and theoretical results, as shown in figure \ref {fig:comparison}. It is seen that, for relatively small $\beta_{AT}<35\degree$, the present theory compares remarkably well with experimental \cite{hornung1982transition, Mouton2008Experiments} and numerical \cite{vuillon1995reconsideration} results. For relatively large $\beta_{AT} > 36\degree$, it follows very well with experimental \cite{Mouton2008Experiments} and numerical \cite{vuillon1995reconsideration, Mouton2007Mach} results. When $35\degree<\beta_{AT}<36\degree$, it is slightly higher than the experimental \cite{Mouton2008Experiments} and numerical \cite{Mouton2007Mach} results. In general, the present theory is reasonable and valid.
\par In this letter, the essence of hysteresis in the shock wave reflection is revealed.  Since the dissipation of kinetic energy is demonstrated as the action of the equation governing the flow field, of which minimal values correspond to steady states of the system, the hysteresis loop is in fact the projection of valley lines in the dissipation landscape, where the saddle-node bifuractions, i.e. intersection points of valley and ridge lines, are actually the transition points. Therefore, the emergence and disappearance evolution of the dissipation barriers, manifested by the ridge line of the dissipation landscape, is the origin of the reflection hysteresis. The present theory, based on the surface geometry of the action, may be generalized to other hysteresis systems.
\begin{acknowledgments}
We are grateful to professor Xin-Liang Li and You-Sheng Zhang for their helpful discussions.
\end{acknowledgments}

\bibliographystyle{apsrev4-1}
\bibliography{reference}

\end{document}